# Interspecific allometric scaling of unicellular organisms as an evolutionary process of food chain creation

Yuri K. Shestopaloff

## Abstract

Metabolism of living organisms is a foundation of life. The metabolic rate (energy production per unit time) increases slower than organisms' mass. When this phenomenon is considered across different species, it is called *interspecific* allometric scaling, whose causes are unknown. We argue that the cause of interspecific allometric scaling is the total effect of physiological and adaptation mechanisms inherent to organisms composing a food chain. Together, the workings of these mechanisms are united by a primary goal of any living creature - its successful reproduction. This primary necessity of each organism and of the entire food chain is that common denominator, to which all organisms adjust their metabolic rates. In this article, we consider unicellular organisms, while the second paper studies multicellular organisms and the entire concept in more detail. Here, using the proposed concepts and experimentally verified growth models of five different unicellular organisms, we obtain close to experimental findings values of allometric exponents of 0.757 for the end of growth and 0.853 for the beginning of growth. These results comply with experimental observations and prove our theory that the requirement of successful reproduction within the food chain is an important factor shaping interspecific allometric scaling.

**Introduction**

In order to support their life cycle, living organisms have to produce energy. The rate of energy production is called metabolic rate (denoted below as *B*). Metabolic rate increases slower than the total mass *M* of organisms [1-5]. This phenomenon is called metabolic allometric scaling. Mathematically, it is described as

$B = aM^b$ (1)

where *a* is a constant; *b* is the allometric exponent.

Two different types of allometric scaling are considered. Intraspecific allometric scaling relates rather to ontogenetic development, while the interspecific scaling describes this effect across different taxa. In [6], an explanation of *intraspecific* allometric scaling was proposed and proofs were provided. However, the fundamental causes of *interspecific* allometric scaling are still unknown [7-12]. Presently, consensus is that this phenomenon rather depends on different interacting factors acting at different organizational levels [11-13].

In terms of generality, the causes of this phenomenon should be on par with its generality, very likely embracing *all* living species, from the smallest microbes to the biggest animals. Such fundamental causes are unlikely to be constraints imposed by a particular physiological mechanism, given the wide range of adaptation capabilities of living organisms and environmental conditions they can thrive in. Living organisms can "overwrite" constraints imposed by particular mechanisms. In [14], the author says: "Although organisms cannot obviate the effects of physical laws and processes, the consequences of these effects can be altered by ontogenetic or phylogenetic alternations in geometry, shape, or orientation as well as in body size". In [8], the authors express a similar opinion about to the role of vascular networks: "The vascular supply network is adapted to the needs of the cells at their working limit. We conjecture that the optimization of the arterial tree by fractal design is the result rather than the cause of the evolution of metabolic rate scaling."

We argue that it is the entire evolutionary process of organic life development (based on numerous physiological mechanisms), which created and presently reproduces the living world as a continuous food chain, is largely responsible for interspecific allometric scaling and its stability. This dynamic balancing and rebalancing of the whole food chain and its parts is based on at least two principles:

(a) The food chain cannot be broken (in any case, for long); (b) Biochemical mechanisms, bio-mechanical constraints and different physical "denominators" (meaning parameters, characterizing interaction of different taxa in the food chain) were evolutionarily developed in such a way that they allow organisms to adapt to a very wide spectrum of different environmental conditions, far exceeding constraints imposed by particular physiological mechanisms.

## Methods

The answer to the puzzle of interspecific metabolic allometric scaling has been sought in three major groups of factors: (a) *biochemical* mechanisms responsible for the energy production and other biological functions supporting the organism's existence; (b) different *environmental* factors affecting organisms' development, like temperature, type of nutrition, etc. (c) *bio-mechanical* constraints, like buckling of bones or trees' trunks, or mechanical pressure the bones can sustain, etc.

The leading factor in organisms' development is the environment, whose characteristics living organisms must fit in order to survive. A similar idea was expressed in [14] as follows: "… it is possible to view organic evolution as an extended 'experiment' in how organisms respond to and cope with the laws governing chemical and physical phenomena." Biochemical machinery and bio-mechanical constraints *serve* the purpose of adaptation, so that the living organisms must exercise the maximum possible flexibility available within these two groups of factors. Environment is a scene; living organisms are actors trying to remain on the scene as long as possible - these are the rules of organic life. As the author of [14] says, "evolution is constrained by physical laws, but … the effects of these laws can be modified by biological innovation".

From this arrangement, the following methodological approach to the problem of interspecific allometric scaling follows:

(a) Determine the factors, which are common across all taxa and are critical for the creation, organization, evolutionary development and existence of the living world and for the creation of a food chain in particular;

(b) Understand the overall structure, dynamics and interrelationships of different factors, together defining interspecific allometric scaling.

## Results

**Variability of metabolic characteristics**

Our previous studies explored intrinsic biochemical and physiological mechanisms, which could impose universal constraints enforcing interspecific allometric scaling. However, these studies rather confirmed that the adaptation capabilities of real organisms allow organisms to adapt to a wide range of environmental conditions, so that it is unlikely that one or a group of particular mechanisms could define the intrinsic nature of allometric scaling.

Let us consider metabolic characteristics of unicellular organisms, assuming that the nutrient consumption strongly correlates with a metabolic rate. Experimental data are from [15,16] for *Schizosaccharomyces pombe*; from [17] for *B. subtilis*, from [18] for *Escherichia coli*, and from [19] for *Amoeba*. We use a growth model validated by these and other experimental data, and find nutrient consumption for whole organisms and per unit of volume. For *Staphylococcus*, we do not have experimental data, so we use the growth model alone. For validation of obtained results, we also use experimental results for excised cells growing in *vitro* in culture [20,21].

**Metabolic rates of S. pombe**

In [22-25], a method for finding growth and metabolic characteristics of microorganisms was introduced and verified by experimental data for *S. pombe*, *Amoeba*, *Schizosaccharomyces cerevisiae*. A similar method was successfully used for the study of growing livers and liver transplants [26,27]. The basis of this method is the fact, cross-verified by different experimental data and observed effects, that the amount of nutrients, used for biomass production at each moment, represents a certain fraction of the total consumed nutrients. Thus, once we know from experiments how the mass of a growing organism (or its constituents) increases, we can find the total amount of consumed nutrients. Detailed consideration of this method for *S. pombe* is in [25]. All growth models and results used in this study are considered in detail in work [28].

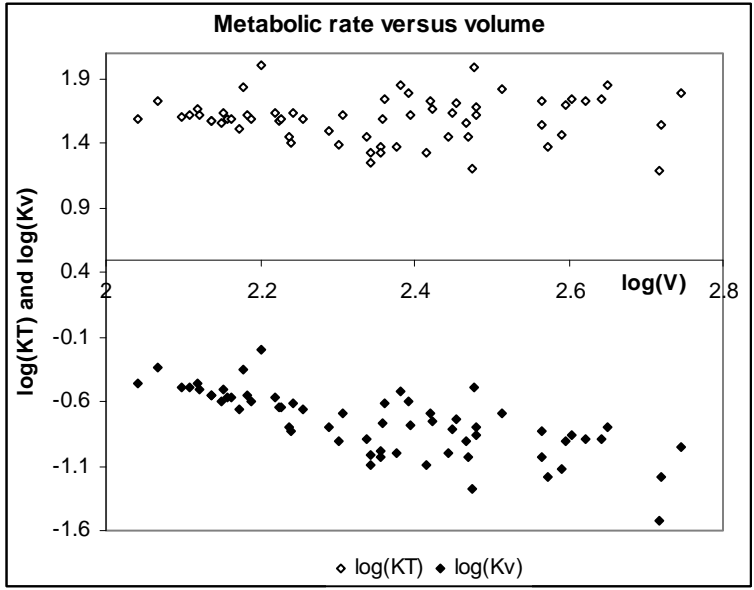

Fig. 1. Metabolic rate versus volume, in logarithmic scale. Black diamonds denote metabolic rate per unit of volume ($K_V$); diamonds denote the total metabolic rate ($K_T$).

Fig. 1 shows the nutrient influx (amount of nutrients per unit time) for *S. pombe* at the end of growth, obtained this way from experimental observations done in [15]. The metabolic rate per unit volume decreases with the increase of volume (for the constant density, we can substitute volume for mass). The trend for the metabolic rate per entire organism changes little with mass increase (the upper ensemble). Experimental data for excised cells growing in cultures [20,21] for mammalian hepatocytes, dermal fibroblasts, skeletal myoblasts, and avian dermal fibroblasts, and their discussion in [29], show that except for the weak allometric scaling for hepatocytes, the rest of cells shows little allometric scaling depending on the body size in the range of masses of several orders of magnitude. We can be certain that sizes of studied cells were different too, and the above conclusion about weak allometric scaling is valid with regard to the cell size too. Such a characteristic behavior we observe for our data on Fig. 1 for the metabolic rate of whole unicellular organisms.

Therefore, the obtained result for *S. pombe* complies with experimental observations for other single cells, which adds credibility to our approach for finding metabolic properties of single cells.

**Metabolic characteristics of unicellular organisms**

Similarly, we can find metabolic characteristics for *E. coli* and *B. subtilis*. *Amoeba's* nutrient consumption was found using the *Amoeba's* growth model from [24,25] and experiments from [19]. *Staphylococcus's* growth was modeled by a growing sphere. The detailed considerations is presented in @@@ Results are presented in Table 1 and Fig. 2. All data points for the total nutrient consumption at the end of growth (division phase) are located very close to a regression line (Fig. 2A). Nutrient consumption per unit volume exhibits high diversion, with the overall decreasing trend (Fig. 2B).

Table 1. Metabolic properties of unicellular organisms. Nutrient consumption per unit of volume ($k_{V\max}$) and the total maximal nutrient consumption ($K_{max}$); '*pg*' denotes *picogram*.

| Organism | $V_{max}$, $\mu m^{-3}$ | $k_{V\max}$, $pg \cdot \mu m^{-3} \cdot \min^{-1}$ | $K_{max}$, $pg \cdot \min^{-1}$ |
|---|---|---|---|
| *B. subtilis* | 0.617 | 1.066 | 0.658 |
| *Staphylococcus* | 2.145 | 0.552 | 1.184 |
| *E. coli, 3-Linear* | 3.999 | 2.65 | 1.59 |
| *E. coli, 2-Linear* | 16.975 | 1.71 | 7.34 |
| *S. pombe* | 325.4 | 0.195 | 63.59 |
| *Amoeba* | $1.88 \cdot 10^7$ | 0.013 | $2.44 \cdot 10^5$ |

Allometric exponent *b* can be found using regression lines, or calculating them for pairs of data points and then finding an average value. For two data points corresponding to masses *M* and *m*, and metabolic rates $K_M$ and $K_m$, we have

$$(K_M / K_m) = (M / m)^b \qquad (2)$$

Solving this equation, we find

$$b = \frac{\log(K_M / K_m)}{\log(M / m)} \qquad (3)$$

The value of allometric exponent found this way for the maximal metabolic rate (all pairs include *amoeba* to provide a greater volume range) is $b = 0.757 \pm 0.012$. This value corresponds to the end of growth (the division phase). For the minimal metabolic rate (at the beginning of growth) $b = 0.853 \pm 0.069$.

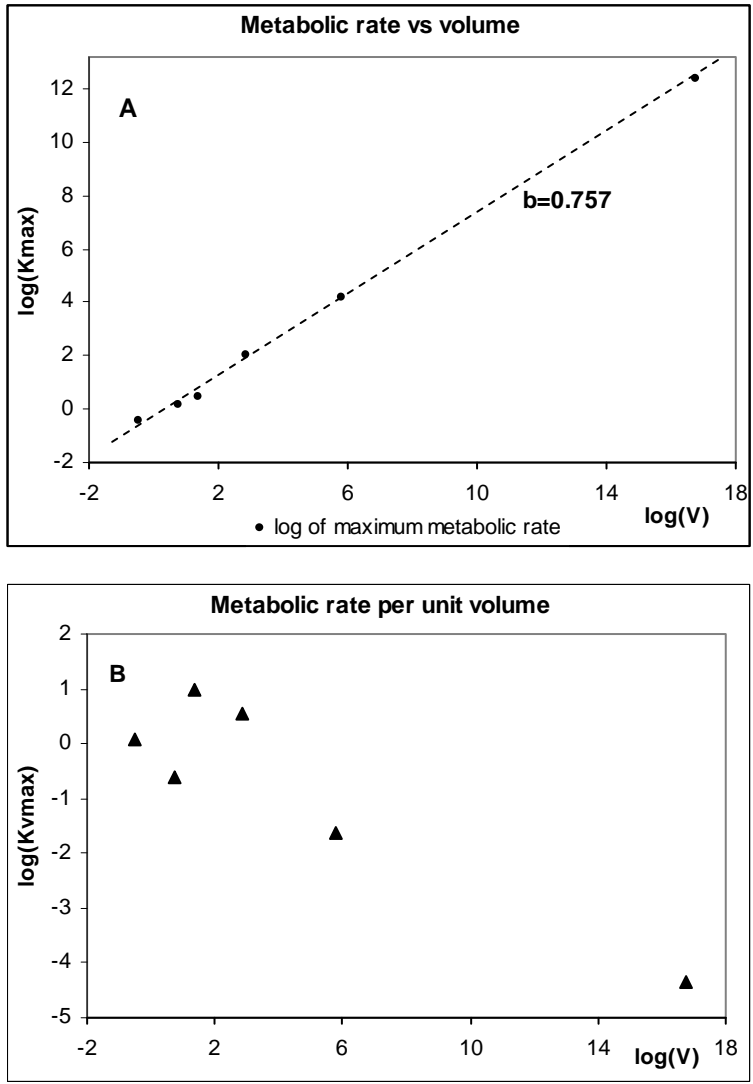

Fig. 2. Change of metabolic rate depending on volume, in logarithmic scale. Data points from left to right correspond to *B. subtilis*, *Staphylococcus,* two data sets for *E. coli*, *S. pombe*, a*moeba*. Experimental data are from [17] for *B. subtilis*, [18] for *E. coli*, [15] for *S. pombe*, [19] for *amoeba*. A - whole organisms; B - per unit volume.

If we exclude *amoeba* and find allometric exponents relative to *S. pombe*, then $b = 0.773 \pm 0.053$ for the maximal metabolic rate (the diversion of data for the minimal metabolic rates in this case is too large to be considered).

Thus, we obtained close values, within 2% of relative difference, although the referenced organisms, *amoeba* and *S. pombe*, are very different organisms and have different size and geometry: *amoeba* was modeled by a disk increasing in two dimensions, whose height remains constant, while *S. pombe* was modeled by an elongating cylinder with a constant diameter. Nonetheless, the actual allometric exponents are very close, which, according to our theory, is the result of evolutionary equalization of metabolic capabilities within the food chain, so that none of the organisms could have significant advantage in order to not compromise the balanced state of the food chain.

**Discussion**

A greater metabolic rate at active phase of growth, compared to the division phase (0.853 versus 0.757), rather should be expected, because of the high rate of biomass synthesis at the growth phase, in addition to other activities. The value of $b = 0.853 \pm 0.069$ complies with known experimental data. It is not far from the experimentally found values of allometric exponent $b = 0.872 \pm 0.029$ for the exercise-induced maximal metabolic rates in mammals, $b=0.849$ for exercising non-athletic animals [8], values $b=0.86$ and $b=0.87$ from review [9] for mammals.

Review [9] shows a wide range of allometric exponents for unicellular organisms, from 0.608 to 1.09. The mostly referenced studies report values close to 0.75, which complies with our results, but only for the maximal metabolic rate. Pytoplankton has an allometric exponent of 0.88, so that our number of 0.853 is not unrealistic.

It would be reasonable to expect a greater dispersion of metabolic rates at the beginning of growth, given the diversity of initial conditions and sizes, than at the end of growth, when organisms switch to a division phase, which is metabolically a well defined and stable process. Our study confirms this consideration.

So, our results support the prediction of the proposed theory that the metabolic rate is rather "equalized" across different unicellular organisms. The need in food acquisition and preservation of a balanced state of the entire food chain adjusts the functioning of main organismal metabolic mechanisms to an average for the food chain regimes, so that all species will be able to obtain sufficient food for successful reproduction, but without jeopardizing the balanced state of the whole food chain.

Living organisms were evolutionarily developing from small forms to larger ones, in one way or another inheriting foundational life building blocks, like mitochondria, ribosome, cells, blood circulation systems, skeletons, etc. Metabolism as such is the foundation of living matter, and so metabolic mechanisms should be propagated through the evolutionary chain too. Given the universality and omnipresence of metabolic allometric scaling, the fundamental properties of metabolism for the same functional regimes should be inherited too, so that the close highest values of allometric scaling for unicellular organisms and exercising mammals should not be of great surprise. The same is true for the close values of 0.757 of allometric exponent for unicellular organisms at a division phase, and the basal metabolic rates (minimal energy requirements at rest) of 0.74 obtained for mammals, because many metabolically active cells in multicellular organisms, in fact, are the grown cells whose division is suppressed.

What kind of environmental factors are most critical for the organisms' survival and reproduction? Apparently, this is a combination of factors. However, once the base factors, like the temperature range, are in place, the primary factors shaping evolutionary development of the living world are the *nutrient acquisition* for successful reproduction, in other words, *competition* for nutrients, and avoiding becoming food too soon to jeopardize the reproduction. And, through fulfilling these requirements, organisms inadvertently keep the entire food chain in a dynamically balanced state.

## Conclusion

We provided proofs of the main conjecture of our study that interspecific allometric scaling is a common creation of organisms composing a food chain. On one hand, organisms have to acquire enough nutrients to support their successful reproduction, while on the other hand the nutrient consumption should not be excessive to jeopardize the balanced state of an entire food chain. These requirements led to "equalization" of the interspecific allometric exponent across different taxa in a food chain. Indeed, our study found a well expressed interspecific allometric regularity for different unicellular organisms. We also calculated values of allometric exponents for the metabolic rates at the beginning of growth, and for the end of growth (the division phase). The results very well comply with available experimental observations.

The observed allometric regularity of the maximal metabolic rate for the whole unicellular organisms proves that factors, responsible for the allometric scaling, affect the *whole* organisms. This result also confirms the main idea of the study that interspecific allometric scaling is defined by adaptation of organisms to environmental factors, within the limits imposed by their belonging to a food chain. Although the primary goal of each organism is its successful reproduction, in a long perspective, it is impossible to fulfill this grand goal without preserving the entire food chain, so that all organisms also participate in its continuous recreation and dynamic balancing and rebalancing.

Unicellular organisms use diverse methods of obtaining nutrients. Many rely on their motility. In the second part of our study, which is the principal one, we consider mammalian species, and, based on their speeds and weight, further prove the validity of the conjecture. We think that a similar study considering mechanical and physical properties can be done for unicellular organisms, but we have no reliable data about these characteristics and environments; besides, they use different modes of motion, which are more difficult to compare from kinematic and energetic perspectives, like swimming of *E. coli* and crawling of *amoeba*.

## Acknowledgements

The author greatly thanks Dr. A. Y. Shestopaloff for fruitful discussions and for the help with editing.

## References

1. Schmidt-Nielsen K. 1984. *Scaling. Why is animal size so important?* Cambridge University Press, New York.
2. Brown JH, West GB. 2000. *Scaling in Biology.* Oxford, UK: Oxford University Press. Brown JH. 2004. The predominance of quarter-power scaling in biology. *Funct. Ecol.*, **18**: 257-282.
3. West GB, Woodruff WH, Brown JH. 2002. Allometric scaling of metabolic rate from molecules and mitochondria to cells and mammals. *Proc. Nat. Acad. Sci.*, **99**: 2473-2478.

4. Savage VM, Gillooly JF, Woodruff WH, West GB, Allen AP, Enquist BJ. Brown JH. 2004. The predominance of quarter-power scaling in biology. *Funct. Ecol.*, **18**: 257-282.

5. Kearney MR, White CR. 2012. Testing metabolic theories. *Am. Nat*., **180**: 546-565.

6. Shestopaloff YK. 2016. Metabolic allometric scaling model. Combining cellular transportation and heat dissipation constraints. *J. Exp. Biol.*, **219**, 2481-2489.

7. White CR, Kearney MR, Matthews PGD, Kooijman SALM, Marshall DJ. 2011. Manipulative test of competing theories for metabolic scaling. *Am. Nat*., **178**: 746-754.

8. Weibel ER, Hoppeler H. 2005. Exercise-induced maximal metabolic rate scales with muscle aerobic capacity. *J. Exp. Biol.,* **208**: 1635-1644.

9. Glazier D. 2005. Beyond the '3/4-power law', variation in the intra-and interspecific scaling of metabolic rate in animals. *Biol. Rev.*, **80:** 611-662.

10. Glazier D. 2010. A unifying explanation for diverse metabolic scaling in animals and plants. *Biol. Rev.*, **85**: 111-138.

11. Glazier D. 2014. Metabolic scaling in complex living systems. *Systems*, **2**: 451-540.

12. Kozlowski J, Konarzewski M. 2004. Is West, Brown and Enquist's model of allometric scaling mathematically correct and biologically relevant? *Funct. Ecol.*, **18**: 283-289.

13. Darveau C-A, Suarez RK, Andrews RD, Hochachka PW. 2002. Allometric cascade as a unifying principle of body mass effects on metabolism. *Nature*, **417**: 166-170.

14. Niklas KJ. 2013. Biophysical and size-dependent perspectives on plant evolution, *Journal of Experimental Botany*. **64**(15): 4817–4827.

15. Baumgartner S, Tolic-Norrelykke IM. 2009. Growth pattern of single fission yeast cells is bilinear and depends on temperature and DNA synthesis. *Biophys. J*, **96**(10): 4336-4347.

16. Sveiczer A, Novak B, Mitchison JM. 1996. The size control of fission yeast revisited. *J. Cell Sci,* **109**: 2947-2957.

17. Godin M, Delgado FF, Son S, Grover WH, Bryan AK, Tzur A, Jorgensen P, Payer K, Grossman AD, Kirschner MW, Manalis SR. 2010. Using buoyant mass to measure the growth of single cells, *Nature Methods*, **7**(5): 387-392.

18. Reshes G, Vanounou S, Fishov I, Feingold M. 2008. Cell Shape Dynamics in Escherichia coli. *Biophys. J*., **94**(January): 251–264.

19. Prescott DM. 1955. Relations between cell growth and cell division. Reduced weight, cell volume, protein content and nuclear volume of *Amoeba* proteus from division to division. *Exp. Cell Res*. **9**: 328-337.

20. McNab BK. 2008. An analysis of the factors that influence the level and scaling of mammalian BMR. *Comp. Biochem. Physiol. A*, **151**: 5-28.

21. McNab BK. 2009. Ecological factors affect the level and scaling of avian BMR. *Comp. Biochem. Physiol. A*, **152**: 22-45.

22. Shestopaloff YK. 2012a. General law of growth and replication, growth equation and its applications, *Biophysical Reviews and Letters*, **7**(1-2): 71-120.

23. Shestopaloff YK. 2012b. Predicting growth and finding biomass production using the general growth mechanism. *Biophysical Reviews and Letters*, **7**(3-4): 177-195.

24. Shestopaloff YK. 2014b. *Growth as a union of form and biochemistry*. AKVY Press, Toronto.

25. Shestopaloff YK. 2015. Why cells grow and divide? General growth mechanism and how it defines cells' growth, reproduction and metabolic properties. *Biophysical Reviews and Letters,* **10**(4): 209-256.

26. Shestopaloff YK, Sbalzarini IF. 2014. A method for modeling growth of organs and transplants based on the general growth law. Application to the liver in dogs and humans. *PLoS ONE*, **9**(6): e99275. doi,10.1371/journal.pone.0099275.

27. Shestopaloff YK. 2014a. Method for finding metabolic properties based on the general growth law. Liver examples. A general framework for biological modeling. *PLoS ONE*, **9**(6): e99836. doi,10.1371/journal.pone.0099836.

28. Shestopaloff YK. 2016. Biophysical growth and reproduction mechanisms of cells and first principles of life origin and development. arXiv:1609.09421 [q-bio.OT], https://arxiv.org/abs/1609.09421

29. Glazier D**.** 2015. Body-mass scaling of metabolic rate: what are the relative roles of cellular versus systemic effects? *Biology*, **4**, 187-199.